\documentclass[a4paper,11pt]{article}
\pdfoutput=1 

\usepackage{jheppub} 

\usepackage[T1]{fontenc} 

\usepackage{amsmath,amsfonts,amssymb,amsthm}
\usepackage{hyperref}
\hypersetup{
    pdfstartview=FitH,
    bookmarksnumbered=true,
    colorlinks,
    breaklinks,
    urlcolor=RoyalBlue,
    linkcolor=RoyalBlue,
    citecolor=NavyBlue
}
\usepackage{graphicx}
\DeclareGraphicsExtensions{.pdf,.png,.jpg,.mps}
\usepackage{booktabs}
\usepackage[dvipsnames]{xcolor}
\usepackage{tikz}
\usepackage{scrextend}
\usepackage{manfnt,pifont}
\usepackage{url}
\usepackage{cleveref}
\usepackage{subfig}
\usepackage[normalem]{ulem}

\usetikzlibrary{decorations.markings}
\usetikzlibrary{arrows.meta}
\usetikzlibrary{calc}
\usetikzlibrary{bending}
\usetikzlibrary{snakes}

\usetikzlibrary{calc,matrix,decorations.pathmorphing,decorations.markings,arrows,positioning,intersections,mindmap,backgrounds}

\usepackage{amsmath,amsthm,amsfonts,amssymb,amscd,mathtools
}
\usepackage{pifont}

\title{\LARGE{Positivity of the Hypergeometric Coon Amplitude}}

\author[]{Bo Wang}
\affiliation[]{Zhejiang Institute of Modern Physics, School of Physics, Zhejiang University, \\Hangzhou, Zhejiang 310058, China }
\affiliation[]{Joint Center for Quanta-to-Cosmos Physics, Zhejiang University,
\\Hangzhou, Zhejiang 310058, China}

\abstract{We utilize a novel method for the partial-wave unitarity recently suggested in \cite{Rigatos:2024beq} to analyse the hypergeometric Coon amplitude. In this approach we use a new type of harmonic numbers as a basis. Owing to the properties of the harmonic numbers this method bypasses lots of difficulties and can be used to derive a clear and unified form for the partial-wave coefficients. This new approach further streamlines the partial-wave unitarity analysis. As an instance, we re-derive the unitarity bounds from Regge trajectory analysis and find the new critical condition of the hypergeometric Coon amplitude. Our new method also benefits the discussion on the various limits of this amplitude. Specifically, we demonstrate the manifest positivity of super string below $d\leq10$ from our new perspective.}

\emailAdd{b\_w@zju.edu.cn}

\date{\color{black}Jan 7, 2023}

\begin{document} 
\maketitle

\newpage
\section{Introduction and summary}

Since the middle of the last century, people have found a method to construct scattering amplitudes based on consistency principles such as unitarity, crossing symmetry and analyticity: the S-matrix bootstrap. The modern S-matrix bootstrap aims to constrain the parameter space of consistent QFTs non-perturbatively. Bounds and kinks in these spaces sometimes represents the special theories. The S-matrix bootstrap has made significant progress in recent years, see \cite{Kruczenski:2022lot} for a detailed review.

String theory can be regarded as a by-product of the S-matrix bootstrap. A monumental string theory result was when Veneziano derived a meromorphic function to describe the scattering of $2\to2$ open-string tachyons \cite{Veneziano:1968yb} in 1968. Veneziano imposed \textit{polynomial residues}, \textit{dual resonance} and \textit{high-energy behaviour} and obtained
\begin{equation}
    \mathcal{A}^{Ven}(s,t)=\frac{\Gamma(-s+m^2)\Gamma(-t+m^2)}{\Gamma(-s-t+2m^2)}\; ,
\end{equation}
where $m$ is the mass of tachyon. We work in conventions where $\alpha^{\prime}=1$ for open string theory.

Shortly after Veneziano's discovery, Coon introduced an additional deformation $q$ to generalize the Veneziano amplitude \cite{Coon:1969yw,Coon:1972qz}. The Coon amplitude has all the properties of the Veneziano, but a different spectrum. 
This non-linear spectrum has logarithmic Regge behavior and an accumulation point, but its  physical interpretation is still absent. There is a study pointing out that the AdS open-string scattering on D-branes has some qualitative similarities with the Coon amplitude \cite{Maldacena:2022ckr}. In an attempt to understand more thoroughly the space of allowed mathematical functions that can potentially describe tree-level $2 \to 2$ scattering, there is renewed interest in constructing amplitudes that have stringy characteristics. The recent study of \cite{Cheung:2023adk} derived non-trivial deformations of the Coon, Veneziano and Virasoro-Shapiro amplitudes. These new objects have been dubbed hypergeometric amplitudes. 

It seems like these new amplitudes are challenging the uniqueness of perturbative string amplitudes \cite{Caron-Huot:2016icg,Cheung:2022mkw,Geiser:2022exp,Cheung:2023uwn}. However, since we lack a physical understanding of these amplitudes that originate from a bottom-up way\footnote{The recently work \cite{Bhardwaj:2023eus} also explored the potential worldsheet origins of the Coon amplitude \cite{Li:2023hce} from a quantum groups perspective. }, we are not sure if they are of physical interest or just mathematical solutions to the S-matrix bootstrap, and so we need impose additional conditions. In order to rule out the possibility that these solutions are merely some mathematical solutions, we impose unitarity and examine its consequences.

Recent studies have managed to constrain a given amplitude from the factorization of higher-point scattering \cite{Arkani-Hamed:2023jwn,Geiser:2023qqq}. However, a lot of lessons can already be learned from analyzing $2\to2$ tree-level quantities. In this paper, we will focus on hypergeometric Coon amplitude that was recently derived from an amplitudes bootstrap in \cite{Cheung:2023adk}. This amplitude is controlled by two parameters $q$ and $r$. We are able to extract important information by imposing unitarity. We constrain the $q$ and $r$ should be bound as
\begin{equation}
     r>-1 ,  \quad  2q^{-r-1}\geq m^2(q-1)+3-q \; .
\end{equation}
We also derive the critical condition 
\begin{equation}
    2 (q_\infty)^{-r_\infty-1}-m^2 (q_\infty-1)-3=0 \; .
\end{equation}

Partial-wave coefficients encode the transition amplitude between different states. The partial-wave unitarity implies that the S-matrix is unitary, and hence is a critical constraint for any scattering amplitude.

Although the idea is very simple, the analysis of the partial-wave coefficients and the implications of unitarity is a highly complicated task. For the simple Veneziano amplitude, there are some analysis from a few years ago \cite{Boels:2014dka,Maity:2021obe,Arkani-Hamed:2022gsa}. Note that, despite this, we still lack a statement of manifest unitarity derived from the string amplitude perspective below the critical dimension of the (super)string. Subsequent works explored some unitarity properties of the Coon amplitude, including some analytic studies and numerical methods \cite{Figueroa:2022onw,Geiser:2022icl,Chakravarty:2022vrp,Bhardwaj:2022lbz,Jepsen:2023sia}. Further studies of positivity of the partial-wave coefficients can be found in \cite{Rigatos:2023asb} for the case of the hypergeometric Veneziano amplitude that was derived in \cite{Cheung:2023adk}. This case includes one deformation parameter, like the Coon amplitude, however when we introduce both the $q$ and $r$ deformations at the same time, the unitarity analysis becomes extremely complicated, and we urgently need a new method to study the partial-wave unitarity.

The motivation of our study is twofold: one is to understand more aspects of the amplitude itself, such as its consistency with unitarity, while the other is to simplify the partial-wave analysis mathematically. With regards to the latter, recently we proposed a novel basis that accommodates the partial-wave unitarity in the case of the Coon amplitude \cite{Rigatos:2024beq}. The method uses the harmonic numbers as a basis and recasts the problem of deriving the partial-wave coefficients into a simple linear-algebra problem. Our interest is to, also, examine if this method is applicable straightforwardly in more complicated amplitudes.

The hypergeometric Coon amplitude approaches various known amplitudes in different limits, such as the Coon ($q=1$), hypergeometric Veneziano ($r=0$) and Veneziano ($q=1$, $r=0$). By using harmonic numbers as our basis, we can express the partial-wave coefficients in these limits in a unified way. We can obtain these limits straightforwardly because we have discovered a new kind of harmonic number, called the cyclotomic $q$-deformed harmonic number and has never appeared in previous literature. When $q=1$, we obtain the cyclotomic harmonic sums \cite{Ablinger:2011te}; when $r=0$, we obtain the multiple harmonic $q$-series \cite{Bradley_2005}; and when $q=1$ and $r=0$, we obtain the ordinary multiple harmonic numbers.

In this paper, we employ harmonic numbers to obtain all partial-wave coefficients of hypergeometric Coon amplitude in a well-organized form. The main point of our analysis is to pack essential information into harmonic numbers and to reduce the analysis into a linear algebra problem. Moreover, we will discuss the partial-wave unitarity and derive some new bounds. The harmonic number facilitates unitarity analysis and we choose hypergeometric Coon amplitude to demonstrate that. We also discuss various limits of the hypergeometric Coon amplitude and their unitarity. Furthermore, certain indications suggest the potential presence of \textit{low spin dominance} \cite{Figueroa:2022onw,Bern:2021ppb} within the hypergeometric Coon amplitude but from partial-wave analysis perspective. Specifically, we hypothesize that the unitarity bounds are determined by the low spin coefficients. We will use the partial-wave low spin dominance to prove the manifest positivity of superstring below $d\leq 10$ and predict the unitarity of Coon amplitude.

The rest of the paper is organized as follows. In \cref{sec:HypergeometricCoon} we review the setup of hypergeometric Coon amplitude and derive partial-wave coefficients using harmonic numbers. Then we preform some unitarity analysis and obtain new bounds in \cref{sec:Unitarityanalysis}. In \cref{sec:Variouslimits}, we consider the various limits of hypergeometric Coon amplitude and show the manifest positivity of Veneziano and Coon. We conclude the main results of this paper and the future direction in \cref{sec:outlook}.

\section{Hypergeometric Coon Amplitude}\label{sec:HypergeometricCoon}

\subsection{Preliminaries}

The hypergeometric Coon amplitude is given by \cite{Cheung:2023adk},
\begin{equation}\label{eq:hypercoon}
    \mathcal{A}(s,t)=\sum_{n=0}^\infty \frac{q^{\tau(\sigma-n)}}{[n-\sigma]_q} \frac{[\tau+n+r]_q![r]_q!}{[\tau + r]_q![n + r]_q!}\; ,
\end{equation}
where
\begin{equation}
    \sigma = 1+(s-m^2)(q-1), \quad \tau = 1+(t-m^2)(q-1)\; .
\end{equation}
The $s$-channel poles of hypergeometric Coon amplitude are located at
\begin{equation}\label{eq:sN}
    s_N=m^2+[N]_q\; ,
\end{equation}
where $[N]_q$ is the $q$-deformed integer
\begin{equation}\label{eq:Nq}
    [N]_q=\frac{1-q^N}{1-q}\; .
\end{equation}
The residue of hypergeometric Coon amplitude on these poles is more complicated, including two deformations of $q$ and $r$. But, as we shall explicitly demonstrate, we can still use the basis of harmonic numbers to perform the partial-wave unitarity analysis. The residue at the $s$-channel poles $s_N$ is given by:
\begin{equation}\label{eq:defresidue}
\mathop{\operatorname{Res}\ \; }\limits_{s=[N]_q} \mathcal{A}(s, t) \equiv \mathcal{R}^{q,r}(t,N)= {q^N} \prod_{n=1}^{N} \left(1+ (t-m^2) \frac{q^{n+r}}{[n+r]_q} \right)\; ,
\end{equation}
where $N=0,1,2,3,\cdots$. In this form, the limits $q\to1$ and $r\to0$ are straightforward to consider. Packaging the residues of Veneziano, Coon and hypergeometric amplitude in a unified form is a key observation of the following analysis.

The hypergeometric Coon amplitude is supported by finite spin on a given fixed pole $s_N$. We can use the Gegenbauer polynomial to decompose the expression for the residue
\begin{equation}\label{eq:residuegegen}
    \mathcal{R}^{q,r}(t,N)=\sum_{\ell=0}^{N}c_{N,\ell}  C^{\left(\frac{d-3}{2}\right)}_\ell\left(1+\frac{2t}{s_N-4m^2}\right)\; ,
\end{equation}
where $d$ is the spacetime dimension. Let us denote $\frac{d-3}{2}$ by $\alpha$ and use the shorthand $\mathcal{N}=s_N-4m^2=[N]_q-3m^2$.

As previously discussed, the lack of an underlying physics theory introduces some uncertainty regarding unitarity. We need to impose the non-trivial condition
\begin{equation}
    c_{N,\ell}\geq 0, \qquad \forall \;  N,\ell
\end{equation}
to ensure the system ghost-free. Even if we temporarily set aside the unitarity analysis, deriving a complete expression for the partial-wave coefficients in a closed-form remains a highly challenging task. One approach involves solving for all coefficients $c_{N,l}$ using the orthogonality relations of Gegenbauer polynomials
\begin{equation}\label{eq:gegen_ortho}
\int_{-1}^{+1} dx C_{\ell}^{(\alpha)} (x) C_{\ell^{\prime}}^{(\alpha)}(x)
(1-x^2)^{\alpha-\frac{1}{2}} = 2 \mathcal{K}(\ell,\alpha) \delta_{\ell \ell^{\prime}}
\;			,
\end{equation}
where $\mathcal{K}(\ell,\alpha)$ is the normalization factor
\begin{equation}\label{eq:normalizationgegen}
\mathcal{K}(\ell,\alpha) = \frac{\pi\Gamma(\ell+2\alpha)}{2^{2\alpha}\ell! (\ell+\alpha) \Gamma^2(\alpha)}
\;      .
\end{equation}
However, this approach culminates in a resulting expression that is intricate and not suggestive at all,
\begin{equation}\label{eq:integralrep}
    c_{N,\ell}= \frac{1}{2 \mathcal{K}(\ell,\alpha)} \int_{-1}^{+1} dx_N (1-x_N^2)^{\alpha-\frac{1}{2}}C_{\ell}^{(\alpha)} (x_N) \mathcal{R}^{q,r}\left((x_N-1)\frac{s_N-4m^2}{2},N \right)\; .
\end{equation}
Notice that this integral expression does not lend itself well to unitarity analysis.

From the integral representation \cref{eq:integralrep} we can use the generating function of the Gegenbauer polynomial in order to derive the partial-wave coefficients as nested sums. This method has been successful for the case of Veneziano, Coon and hypergeometic Veneziano amplitude \cite{Maity:2021obe,Chakravarty:2022vrp,Rigatos:2023asb,Rigatos:2024beq}. An alternative approach is the use of double contour-integrals to represent and solve for the partial-wave coefficients. This method has, also, been very successful in studies of the Veneziano and the Coon \cite{Arkani-Hamed:2022gsa,Bhardwaj:2022lbz}. Note that \cite{Geiser:2022icl} use the contour integral representation in studying generalisations of string amplitudes. Both of these methods are applicable in the case of the hypergeometric Coon amplitude, but since they have limitations we opt to utilize a recently suggested approach \cite{Rigatos:2024beq}. We will use harmonic numbers as a basis and derive the partial-wave coefficients  from a simple linear-algebra problem.

\subsection{Harmonic numbers and Partial-wave coefficients}
We begin by introducing the concept of cyclotomic $q$-deformed harmonic numbers,
\begin{equation}\label{eq:hm}
    Z^{q,r}_{\{i_1,i_2,\cdots,i_k\}}(N) \equiv \sum_{n=1}^{N} \frac{q ^{n+r} }{[n+r]_q^{i_1}}Z^{q,r}_{\{i_2,i_3,\cdots,i_k\}}(n-1)\,  ,
\end{equation}
where $i$ denotes the symbol letters. To establish the necessary groundwork, we also define
$Z^{q,r}_{\{\}}(N)=1$ and $Z^{q,r}_{\{i_1,\cdots,i_k\}}(0)=0$, $\forall\; k\geq 1$. In our specific context, we specialize to the case that every index $i$ is equal to $1$. Consequently, we introduce the notation:
\begin{equation}
    Z^{q,r}_{k}(N)\equiv Z^{q,r}_{\{1_1,,1_2,\cdots,1_k\}}(N)\; ,
\end{equation}
$k$ is the length of the symbol. This notation provides a concise representation for our purposes.
We also  provide a single summation form for this harmonic numbers,
\begin{align}\label{eq:singlehs}
    Z_{ {k}}^{q,r}(N)=\frac{(q-1)^k }{\left(q^{-N-r};q\right){}_{N}} {\sum _{j=0}^{N}\binom{j}{k} \frac{ \left(q^{-N};q\right){}_{N-j}}{(q;q)_{N-j}}} q^{-(N-j)r}\; , \quad
    (x;q)_n\equiv \prod_{i=0}^n(1-x  q^i)\, .
\end{align}
This expression is $r$-analogous version of \cite[Equation 43]{Rigatos:2024beq}.

Now, let us go back to the hypergeometric Coon amplitude. We remind the reader that its residue can be expressed in the form of \cref{eq:defresidue}. Surprisingly, upon closer examination, we discover that its polynomial coefficients align precisely with the harmonic numbers, which follows from the identity
\begin{equation}
     \prod_{n=1}^{N} \left(1+t \frac{q^{n+r}}{[n+r]_q}\right) = \sum_{n=0}^{N} t^{n} Z^{q,r}_{ {n}}(N)\; .
\end{equation}
This key identity will lead us toward our final result. By using the binomial theorem, we can express the residue as follows
\begin{equation}\label{eq:harmonicR}
     \mathcal{R}^{q,r}(t,N)= {q^N} \sum_{n=0}^{N}\sum_{k=0}^{N} \binom{k}{n} (-m^2)^{k-n} t^{n} Z^{q,r}_{ {k}}(N)\; .
\end{equation}

Our purpose is to reduce the full procedure into the linear algebra problem. A straightforward idea is to expand the Gegenbauer polynomial into series of $t$ in \cref{eq:residuegegen}. Using the identity
\begin{equation}\label{eq: ident_gegen}
    C^{(\alpha)}_\ell(x)=
\frac{(2\alpha)_{\ell}~\;}{\Gamma(\ell+1)} 
{}_2F_1\left(-\ell,\ell+2\alpha;\alpha+\frac{1}{2};\frac{1-x}{2}\right)
\; ,
\end{equation}
we obtain
\begin{align}\label{eq:expandR}
   \mathop{\operatorname{Res}\ \; }\limits_{s=[N]_q} \mathcal{A}(s, t) & =\sum_{\ell=0}^{N} c_{N, \ell} C_\ell^{\left(\alpha\right)}\left(1+\frac{2 t}{[N]_q-3 m^2}\right) \nonumber \\
    &=\sum_{n=0}^{N} \sum_{\ell=0}^{N} c_{N,\ell} \underbrace{ \frac{\Gamma (\ell+2\alpha) (-\ell)_n (\ell+2\alpha)_n}{n!\, \ell! \, \Gamma (2\alpha) \left(\frac{1}{2}+\alpha\right)_n} }_{\mathcal{T}_{n,\ell}}  \left(-\frac{t}{\mathcal{N}}\right )^n \;.
\end{align}
where $\mathcal{N}=[N]_q-3m^2$.

Up to this point, we have obtained two distinct expansions for the residue $\mathcal{R}^{q,r}(t,N)$. The first expansion \cref{eq:harmonicR} is based on the harmonic numbers and the second one \cref{eq:expandR} involves the decomposition of the partial-wave coefficients. By equating the two expressions and considering that they are both series expansions in $t$, we can formulate the sum-rules equations in the following way:
\begin{equation}
    \sum_{\ell=0}^{N} c_{N,\ell} \mathcal{T}_{n,\ell} \left(-\frac{1}{\mathcal{N}}\right )^n ={q^N} \sum_{k=0}^{N} \binom{k}{n} (-m^2)^{k-n}  Z^{q,r}_{ {k}}(N)\; .
\end{equation}
As a result, the problem elegantly simplifies into an elementary linear algebra exercise with its solution being written as
\begin{align}\label{eq:res_doublehm}
     c_{N,\ell} ={q^N} \sum_{n,k=0}^{N} \underbrace{\binom{n}{\ell} \frac{\sqrt{\pi} }{\mathcal{K}(\ell,\alpha)} 
     \frac{(-1)^{\ell} (\alpha)_{\frac{1}{2}+n}}{(\ell+2\alpha)_{1+n}}  }_{\mathcal{T}_{n,\ell}^{-1}}  \underbrace{\binom{k}{n} (-m^2)^{k-n}}_{\text{external mass}} \underbrace{ {\left(3 m^2-[N]_q\right)^n} \vphantom{\binom{a}{n}}}_{\text{scattering angle}} Z^{q,r}_{ {k}}(N)\; .
\end{align}

Readers may discern that this result bears a striking resemblance to the partial-wave coefficients of the Coon amplitude we derived earlier \cite{Rigatos:2024beq}. The only distinction lies in the variation of harmonic numbers. This aligns with our anticipations. Let us revisit our commentary on this structure. Each constituent of this structure is associated with a specific contribution. The inverse matrix $\mathcal{T}_{n,\ell}^{-1}$, external mass and scattering angle are universal. The final part is a basis of harmonic numbers. Comparing with the case of the Coon amplitude, we only replace the
$q$-deformed harmonic number by cyclotomic $q$-deformed harmonic numbers $Z^{q,r}_k(N)$.

Before we proceed, we stress that the summation over $n$ in \cref{eq:res_doublehm} can be done analytically. We have the liberty to extend the upper bound of $n$ to infinity due to the binomial coefficient. Consequently, we obtain a single-sum
\begin{align}\label{eq:res_singlehm}
    c&  _{N,\ell}=q^{N} \sum_{k=0}^{N}2 (-1)^{\ell } \Gamma (2 \alpha ) (-m^2)^k  (\alpha +\ell ) Z_{ {k}}^{q,r}(N)   \, _3\tilde{F}_2\left(1,\alpha +\frac{1}{2},-k;1-\ell ,\ell +2 \alpha +1;-\frac{\mathcal{N}}{m^2}\right)
\;      .
\end{align}
where the regularised hypergeometric function ${}_{p}\tilde{F}_{q}(a_1,\ldots,a_p;b_1,\ldots,b_q;z)$ reads
\begin{equation}\label{eq:defreghyp}
{}_{p}\tilde{F}_{q}(a_1,a_2,\ldots,a_p;b_1,b_2,\ldots,b_q;z)
= \frac{{}_{p}F_{q}(a_1,a_2,\ldots,a_p;b_1,b_2,\ldots,b_q;z)}{\Gamma(b_1)\Gamma(b_2)\ldots\Gamma(b_q)}
\;	.
\end{equation}

Let us make an additional comment here. The result \cref{eq:res_doublehm}, imbued with a strong implication, provides us with some insights: for partial-wave coefficients, some aspects are universal, while others are controlled by different harmonic numbers. From a physical perspective, the form of these harmonic numbers is entirely determined by the polynomial residue. We can regard the residue, such as in \cref{eq:defresidue}, as a generating function for a harmonic number. This understanding can even be extended to the Virasoro-Shapiro amplitude \cite{Virasoro:1969me,Shapiro:1970gy}, its hypergeometric generalisation \cite{Cheung:2023adk} and its AdS version \cite{Alday:2022uxp,Alday:2022xwz}.
~\\

\subsection{Matching with previous results}
In this subsection, we make some consistency checks. Our result \cref{eq:res_doublehm} matches \cite[Footnote 30]{Cheung:2023adk} up to an overall factor $q^N$. Choosing $r=0$, we also check our results against the Coon coefficients \cite[Equation 67]{Cheung:2022mkw} and \cite[Equation 27]{Rigatos:2024beq}. It agrees with hypergeometric amplitude coefficients \cite[Equation 4.13]{Rigatos:2023asb} precisely when $q=1$ and $m^2=-1$.

Unquestionably, our result \cref{eq:res_doublehm} will return to Veneziano when $q=1$, $r=0$. However, there is a small difference in conventions so we require a shift for the quantum number $N$, see \cite[Section 2.3.1]{Chakravarty:2022vrp}. We will present the explicit expression in \cref{sec:manifestpos}.
~\\

\section{Unitarity analysis}\label{sec:Unitarityanalysis}
In this section, we concentrate on the partial-wave unitarity of the hypergeometric Coon amplitude. We will fist present some low-lying coefficients and directly derive some new unitarity bounds. Then, using harmonic numbers, we investigate the behavior of the Regge trajectories. We will present some leading and sub-leading results and add some discussion and comments. Finally, we consider the low-spin scenario to demonstrate the effectiveness of harmonic numbers.

\subsection{Straightforward unitarity bound}\label{sec:straightunitarity}
\begin{table}[h!]
\begin{center}
\begin{tabular}{c|c c c}
 Hyper-Coon & $\ell=0$ & $\ell=1$ & $\ell=2$  \\ \hline
$N=0$ & $Z^{q,r}_0(0)$ &  &  \\
$N=1$ & $\left(\frac{1}{2} (m^2-1) Z^{q,r}_1(1)+Z^{q,r}_0(1)\right)\; {\color{red}{q}} $  & $ \frac{[1]_q-3 m^2}{2 (d-3)} Z^{q,r}_1(1) \;{\color{red}{q}}$ &   \\
$N=2$ & $a_{2,0}\; {\color{red}{q^2}}$ & $a_{2,1}\; {\color{red}{q^2}}$ &  $ \frac{([2]_q-3 m^2)^2 }{2 (d-3) (d-1)} Z_{2}^{q,r}(2)\; {\color{red}{q^2}}$
\end{tabular}
\end{center}
\caption{Partial-wave coefficients of hypergeometric Coon amplitude. All the leading Regge coefficients are manifestly positive when $d\geq4$ and $r>-1$. The coefficients $a_{2,0}$ and $a_{2,1}$ is too heavy in the table hence we list them in \cref{a2}.}
\label{Table:hypercoon}
\end{table}

For the convenience of the readers, we have tabulated some low-lying coefficients in \cref{Table:hypercoon}. We also defined
\begin{subequations}\label{a2}
\begin{align}
    a_{2,0} &= \frac{ (d+8)m^2-2(d+2)m^2 [2]_q+d[2]_q^2 }{4 (d-1)} Z^{q,r}_2(2)+\frac{m^2-[2]_q}{2} Z^{q,r}_1(2) +Z^{q,r}_0(2) \; , \\
    a_{2,1} &= \frac{([2]_q-3 m^2) }{2 (d-3)}\left(Z^{q,r}_2(2) (m^2-[2]_q])+Z^{q,r}_1(2)\right) \; .
\end{align}
\end{subequations}

The highlighted part of $q$ are our distinction from the paper \cite[Footnote 30]{Cheung:2023adk}, which is merely some overall factor $q^N$. Starting from these coefficients, we can make some basic observations: 
\begin{itemize}
    \item It is imperative to ascertain that the harmonic number remains non-negative, as this condition ensures that the leading trajectory matches the Coon amplitude behavior when $r\to0$.
    \item The form of the leading trajectory looks to be very simple. The unitarity bound can be derived directly.
    \item In the hypergeometric Coon amplitude, the subleading trajectory is not obviously positive, and we need a more detailed analysis.
\end{itemize}
After having these three observations, let us start the straightforward unitarity analysis.
~\\

\noindent{\underline{$c_{0,0}$, $c_{1,1}$ and $c_{2,2}$}}
~\\

\noindent These three coefficients provide very sufficient unitarity bounds, which are
\begin{equation}
     r>-1, \quad 1-3m^2\geq 0 \; .
\end{equation}
The first condition ensures the positivity of the harmonic numbers by definition \cref{eq:hm} and the last one constraints the external mass. Note that $r > -1$ is also derived directly from the hypergeometric Veneziano. It is a well-known fact that the Veneziano amplitude has no ghosts when $m^2<-1$. Bearing this in mind, we set
\begin{equation}
    -1\leq m^2\leq \frac{1}{3}
\end{equation}
as our parameter range. The parameter range coincides precisely with that of the Coon amplitude as we expect. This boundary is introduced by $c_{1,1}$, and due to the manifest positivity of the harmonic number, the constraints imposed on both the hypergeometric Coon and Coon are fundamentally identical. 
~\\

\noindent{\underline{$c_{1,0}$}}
~\\

\noindent Next, we study the constraint from $c_{1,0}$. The explicit expression can be read
\begin{equation}
    c_{1,0}=\frac{q}{q^{-r-1}-1} \left[2q^{-r-1}- (3-q+m^2 (q-1)) \right]\; .
\end{equation}
Indeed, we have re-derived the bound given by \cite{Cheung:2023adk}
\begin{equation}\label{eq:c10}
    2q^{-r-1}\geq 3-q+m^2(q-1).
\end{equation}
This bound reduces to 
\begin{equation}\label{eq:conditionc10r}
    r\geq -\frac{1+m^2}{2}
\end{equation}
when $q\to 1$. This outcome falls within the range where $r>-1$ and precisely matches the bound $r\geq0$ when $m^2=-1$ \cite{Rigatos:2023asb}. Furthermore, we can also consider the limit as $r\to0$, the partial-wave coefficients that have been proven to be non-negative in the Coon \cite{Bhardwaj:2022lbz,Rigatos:2024beq}.~\\

\noindent{\underline{$c_{2,1}$}}
~\\

\noindent Having dealt with the simple cases, we move on to some more complicated examples to analyze and we next turn our attention to the partial-wave coefficient $c_{2,1}$. Its explicit result is as follows,
\begin{equation}
    c_{2,1}=\frac{([2]_q-3 m^2) }{2 (d-3)} \frac{(1-q) q^{r+3}}{\left(1-q^{r+1}\right) \left(1-q^{r+2} \right) } \left[\left(m^2 (1-q)+q^2-3\right) q^{r+2}+q+1 \right ]\; .
\end{equation}
After suppressing an overall non-negative factor, we are able to derive the unitarity condition
\begin{equation}\label{eq:c21}
    q^{-r-2} (q+1)\geq {3-q^2+m^2 (q-1)}\; .
\end{equation}
Here we point out that \cref{eq:c10} is fully encompasses the range of \cref{eq:c21}. By performing a straightforward scaling, we obtain
$\left(q+1\right) q^{-r-2}\geq 2q^{-r-1}\geq 3-q+m^2(q-1)\geq 3-q^2+m^2(q-1)$. The \cref{eq:c21} can be directly derived from \cref{eq:c10}, which we also observed for the data of the further sub-leading trajectories. Consequently, further discussion of the unitarity bound given by $c_{2,1}$ and additional sub-leading trajectories is immaterial. This result seems as a first hint of the \textit{partial wave low spin dominance}. We note that this claim might seem a bit too strong at first sight, but it will make much more sense when we discuss explicitly low-spin dominance in \cref{sec:lowspin}. Within the same Regge trajectory, it appears that the unitarity bound is entirely controlled by the low spin partial-wave coefficient. While this remains a preliminary observation, we will revisit it in subsequent discussions.
~\\

\noindent{\underline{$c_{2,0}$}}
~\\

\noindent Now, we will focus on the $c_{2,0}$ example that is much less under control. We first present its complete expression and its unitarity condition, 
\begin{align}
    c_{2,0}=& \frac{q^2}{4 (d-1) \left(1-q^{r+1}\right) \left(1-q^{r+2}\right)} \left.\Big [ 2 (d-1) (q+1) \left(m^2(1-q)+q^2-3\right) q^{r+1}+4 (d-1) \right. \nonumber \\
    & \left. +q^{2 r+3} \left(d \left(m^2(1-q)+q^2-3\right)^2+4 (m^2 (q-1)-1) \left(2 m^2 (q-1)-q^2+2\right)\right) \right] \geq0\; .
\end{align}
Extracting information about partial-wave unitarity from this equation seems rather cumbersome, hence we do not intend to do it here. However, in \cref{sec:lowspin}, we will utilize the properties of the harmonic number to benefit the low spin analysis. Furthermore, considering $m^2=-1$, this value can be pushed to $d=10$. Therefore, we do not discuss the unitarity bound of $c_{2,0}$ here, but contemplate some special limits. Considering $q\to 1$, we find a decomposition
\begin{equation}
    c_{2,0}\Big|_{q\to1}=\frac{1}{(r+1) (r+2)}\left(r-\frac{\eta(d,m)-\delta(d,m)}{2 (d-1)}\right) \left(r-\frac{\eta(d,m)+\delta(d,m)}{2 (d-1)}\right)\; ,
\end{equation}
where
\begin{align}
    \eta(d,m)&=(1-d)(1+m^2)\; , \\ 
    \delta(d,m)&=\sqrt{(d-1) \left(d-9 m^4+12 m^2-5\right)}\; .
\end{align}
The interesting aspect of this form is that when $m^2=-1$, the factorization reduces to
\begin{equation}
    \frac{1}{(r+1) (r+2)}\left(r-\frac{\sqrt{d-26}}{\sqrt{4d-4}}\right) \left(r+\frac{\sqrt{d-26}}{\sqrt{4d-4}}\right)=\frac{1}{(r+1) (r+2)}\left( r^2+\frac{26-d}{4 (d-1)} \right)
\end{equation}
and we directly observe the critical dimension $d=26$ of the bosonic string. This relation has also been observed in \cite{Boels:2014dka}.
~\\

\noindent{\underline{Critical condition of $q$ and $r$}}~\\

\noindent Inspired by the low spin dominance, let us explore the critical condition of $q$ and $r$. The partial-wave coefficients is always positive once the critical condition has been satisfied. Then the hypergeometric Coon amplitude is ghost-free in all dimensions. We begin by considering the asymptotic behavior of $c_{N,0}$ at $N$ approaches infinity,
\begin{align}
    \lim_{N\to\infty }c_{N,0}&\propto \lim_{N\to\infty } \left[  q^N-\frac{N}{2}  q^N \left(-3 m^2 +[N]_q\right)  Z^{q,r}_1(N) +\cdots \right] \nonumber \\
    &\propto \lim_{N\to\infty } \left[ 2 q^{-r-1}-[N]_q \left(m^2 (q-1)-q^{N}+3\right) \right] \nonumber \\
    &=2 q^{-r-1}-m^2 (q-1)-3\; .
\end{align}
In the second step we utilize the explicit expression of harmonic numbers \cref{eq:singlehs}. Finally, the critical condition involving parameters $q$ and $r$ can be expressed as
\begin{equation}\label{eq:criticalcondition}
    2 (q_\infty)^{-r_\infty-1}-m^2 (q_\infty-1)-3=0 \; .
\end{equation}
Readers can check that our approach aligns with the method described by \cite{Figueroa:2022onw}, yielding the same result. For completeness and clarity, we review the main steps. First of all, we expand the residue
\begin{equation}
    \mathcal{R}^{q,r}(t,N)= (b_N)^N \prod_{j=0}^{N-1} (x-x_{j,N})= (b_N)^N \sum_{k=0}^N p_{N,k} x^k\, . 
\end{equation}
where $b_N$ is an overall positive number. The orthogonal relation of Gegenbauer polynomials tells us that the positivity of partial-wave coefficients can be derived from the positivity of $p_{N,k}$, which is related to its roots $x_{j,N}$ being negative. The expression of $x_{j,N}$ is
\begin{equation}
    x_{j,N}=\frac{-2 q^{j-N-r}+m^2 (q-1)-q^{N}+3}{3 m^2 (q-1)-q^{N}+1} \, .
\end{equation}
Note that $x_{j,N}<x_{j+1,N}$ is sufficient to ensure that $x_{N-1,N}$ is negative when $N\to \infty$ ($x_{N-1,N}$ increases monotonically), which means $x_{N-1,N}= 0$ gives the critical condition when $N\to \infty$. This result precisely agrees with the critical condition we derived using harmonic numbers.

In the limit of $r_\infty\to 0$, we re-derive the critical value of $q$ \cite{Figueroa:2022onw,Rigatos:2024beq}
\begin{equation}
    q_{\infty}=\frac{m^2-3+\sqrt{9+2m^2+m^4}}{2m^2}\; .
\end{equation}
Furthermore, when $q_\infty \to 1$, the critical condition \cref{eq:criticalcondition} simplifies to $1$, leading us to conclude that the parameter $r$ exhibits no critical behavior.
~\\

\subsection{Regge trajectory analysis}
From the analysis in the previous subsection, we discerned that directly extracting partial-wave unitarity is technically intricate, which is the reason why we suggested the harmonic number basis. Employing this basis streamlines the analysis of Regge trajectories. We will provide the explicit expression of the leading and sub-leading trajectories.

First, let’s focus on the leading trajectory. By setting $\ell=N$, we obtain the following expression
\begin{equation}\label{eq:leadingtrajectory}
    c_{N,N}=\left(\frac{q}{4}\right)^N\frac{N!}{(\alpha )_{N}}   \left([N]_q-3 m^2\right)^{N} Z^{q,r}_{N}(N)\; .
\end{equation}
We claim that this expression re-produce the unitarity bound we previously derived in \cref{sec:straightunitarity},
\begin{equation}\label{eq:unitaritylead}
     r>-1, \quad 1-3m^2\geq 0 \; .
\end{equation}

In contrast to the Coon amplitude, here the sub-leading trajectory is not always non-negative. However, it introduces new unitarity bounds. Let us begin by expressing it using harmonic numbers:
\begin{equation}
    c_{N,N-1}=\left(\frac{q}{4}\right)^{N} \frac{(N-1)!}{(\alpha )_{N-1}}  ([N]_q-3 m^2)^{N-1} \left(2 Z^{q,r}_{N-1}(N)-N  ([N]_q-m^2) Z^{q,r}_{N}(N)\right)\; ,
\end{equation}
where $N=1,2,3,\cdots$. These expressions yield the non-trivial unitary bounds
\begin{equation}
    2 Z^{q,r}_{N-1}(N)-N  ([N]_q-m^2) Z^{q,r}_{N}(N)\geq0\; ,\quad  \forall\; N \; .
\end{equation}
By substituting the specific expression for harmonic numbers \cref{eq:singlehs}, we arrive at a more intuitive form
\begin{equation}
    2 [N]_q q^{-N-r}-N \left(m^2 (q-1)-q^{N}+3\right)\geq 0 ,\quad  \forall\; N \; .
\end{equation}
As previously discussed in \cref{sec:straightunitarity} , we can once again employ inequality techniques to establish an effective bound. Specifically, we consider the following expression
\begin{equation}\label{eq:unitarysubleading}
    \frac{2}{N}\frac{[N]_q}{q^N} q^{-r} \geq 2q^{-r-1}\geq m^2(q-1)+3-q\geq m^2(q-1)+3-q^N,
\end{equation}
where we use the inequality
\begin{equation}
    \frac{2}{N}\frac{[N]_q}{q^N} q^{-r} \geq 2q^{-r-1} \quad \Longleftrightarrow \quad \frac{1}{N}(1+\frac{1}{q}+\frac{1}{q^2}+\cdots+\frac{1}{q^{N-1}})\geq1\; .
\end{equation}

We conclude that the analysis of the leading trajectory and sub-leading trajectories both give the non-trivial unitarity bounds. Furthermore, upon careful investigation, we discover that these unitarity bounds are primarily determined by the low spin partial-wave coefficients. This observation strengthens the existence of \textit{partial-wave low spin dominance} in the hypergeometric Coon amplitude.

\subsection{Low-spin analysis}\label{sec:lowspin}

As demonstrated in the previous subsections, the low spin partial-wave coefficients play a crucial role in unitarity analysis. The basis of harmonic numbers provides an effective description for these low spin partial-wave coefficients. Let us start from \cref{eq:res_singlehm} and focus on special dimension with low spin. We will discuss the cases where $d=4$ with $\ell=0,1$ and $d=6$ with $\ell=0$. Here we introduce $c^{(d)}_{N,\ell}$ to represent the coefficients in special dimension $d$.

Choosing $d=4$ with $\ell=0$ and substituting the explicit expression of harmonic numbers \cref{eq:singlehs} into \cref{eq:res_singlehm}, we can sum $k$ from $0$ to $\infty$ and the result yields
\begin{align}\label{eq:c4N0}
    c^{(4)}_{N,0}=&\frac{1}{\left(q^{-N-r};q\right){}_{N}} \sum_{j=0}^{N} \frac{\left(q^{-N};q\right){}_{N-j} }{(q;q)_{N-j} }  \frac{q^{-(N-j)r+N}}{(j+1)(1-q)} \nonumber  \\
    & \frac{1}{\mathcal{N}} \left[\left(2 m^2 (q-1)-q^{N}+2\right)^{j+1} - (m^2 (1-q)+1)^{j+1} \right]\; .
\end{align}
Simplifying further using the inequality
\begin{equation}
    \left(2 m^2 (q-1)-q^{N}+2\right) - (m^2 (1-q)+1)= 3 m^2 (q-1)+1-q^{N} \geq 0 \; ,
\end{equation}
where we have leveraged the positivity of the leading trajectory \cref{eq:leadingtrajectory} within this inequality. As a result of our analysis, we can succinctly express the partial-wave coefficients $c^{(4)}_{N,0}$ in a simplified form. 

Next, we still consider $d=4$ but with $\ell=1$.  The resulting partial-wave coefficients are given by
\begin{align}\label{eq:c4N1}
    c^{(4)}_{N,1}=& \frac{-3}{\left(q^{-N-r};q\right){}_{N}} \sum_{j=0}^{N} \frac{\left(q^{-N};q\right){}_{N-j} }{(q;q)_{N-j}} \frac{q^{-(N-j)r+N}}{(1+j)_2(1-q)^2} \nonumber \\
    & \frac{1}{\mathcal{N}^2} \left[(m^2 (1-q)+1)^{j+1} f^{(4)}_1  +\left(2 m^2 (q-1)-q^{N}+2\right)^{j+1} g^{(4)}_1 \right] \; ,
\end{align}
where we define
\begin{align}
    f^{(4)}_1=& (3 j+4) m^2 (q-1)-(j+2) q^{N}+j+4 \; , \\
    g^{(4)}_1=& (3 j+2) m^2 (q-1)+j(1-q^{N})-2 \; ,
\end{align}
Notice that the second line of \cref{eq:c4N1} is always positive due to
\begin{equation}
    f^{(4)}_1-g^{(4)}_1=2 (3+m^2(q-1)-q^N)\geq 0 \; .
\end{equation}

Now, let us delve into a more involved case where $d=6$ with $\ell=1$, the partical-wave coefficients take the form
\begin{align}\label{eq:c6N0}
    c^{(6)}_{N,0}=& \frac{6}{\left(q^{-N-r};q\right){}_{N}} \sum_{j=0}^{N} \frac{\left(q^{-N};q\right){}_{N-j} }{(q;q)_{N-j}} \frac{q^{-(N-j)r+N}}{(1+j)_3 (1-q)^3} \nonumber \\
    &  \frac{1}{\mathcal{N}^3}\left[(m^2 (1-q)+1)^{j+2} f^{(6)}_0  +\left(2 m^2 (q-1)-q^{N}+2\right)^{j+2} g^{(6)}_0 \right] \; ,
\end{align}
where we define
\begin{align}
    f^{(6)}_0=& m^2 (3 n+7) (q-1)-(j+3) q^{N}+j+5\; ,\\
    g^{(6)}_0=& m^2 (3 j+5) (q-1)-j \; .
\end{align}
The second line of \cref{eq:c6N0} is manifestly positive due to $f^{(6)}_0-g^{(6)}_0\geq0$.

\begin{figure}[t]
    \centering
    \subfloat[Unitarity bound from all data]{
        \includegraphics[width=5.8cm]{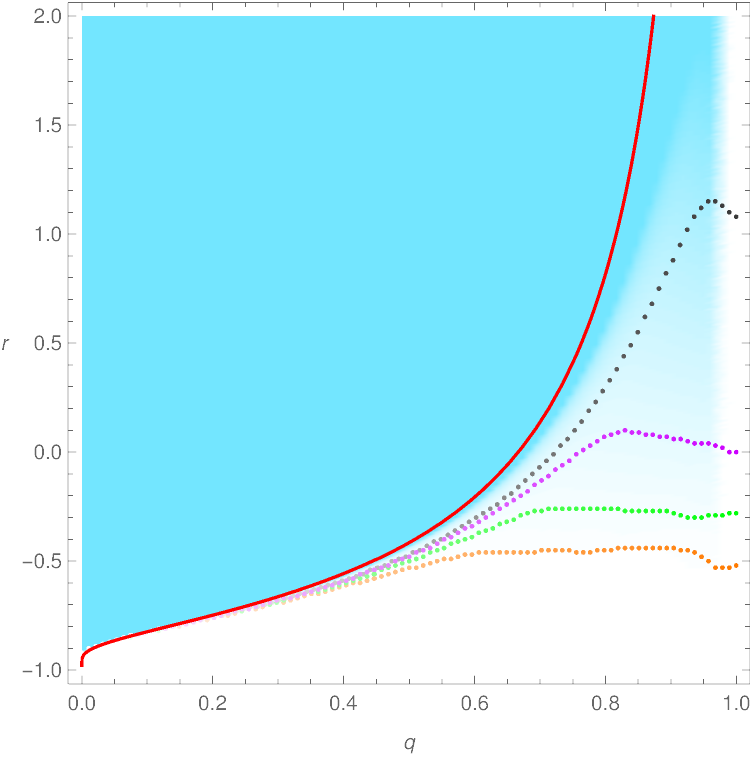}
        } \hfil
    \subfloat[Unitarity bound from spin-0 data]{
        \includegraphics[width=5.8cm]{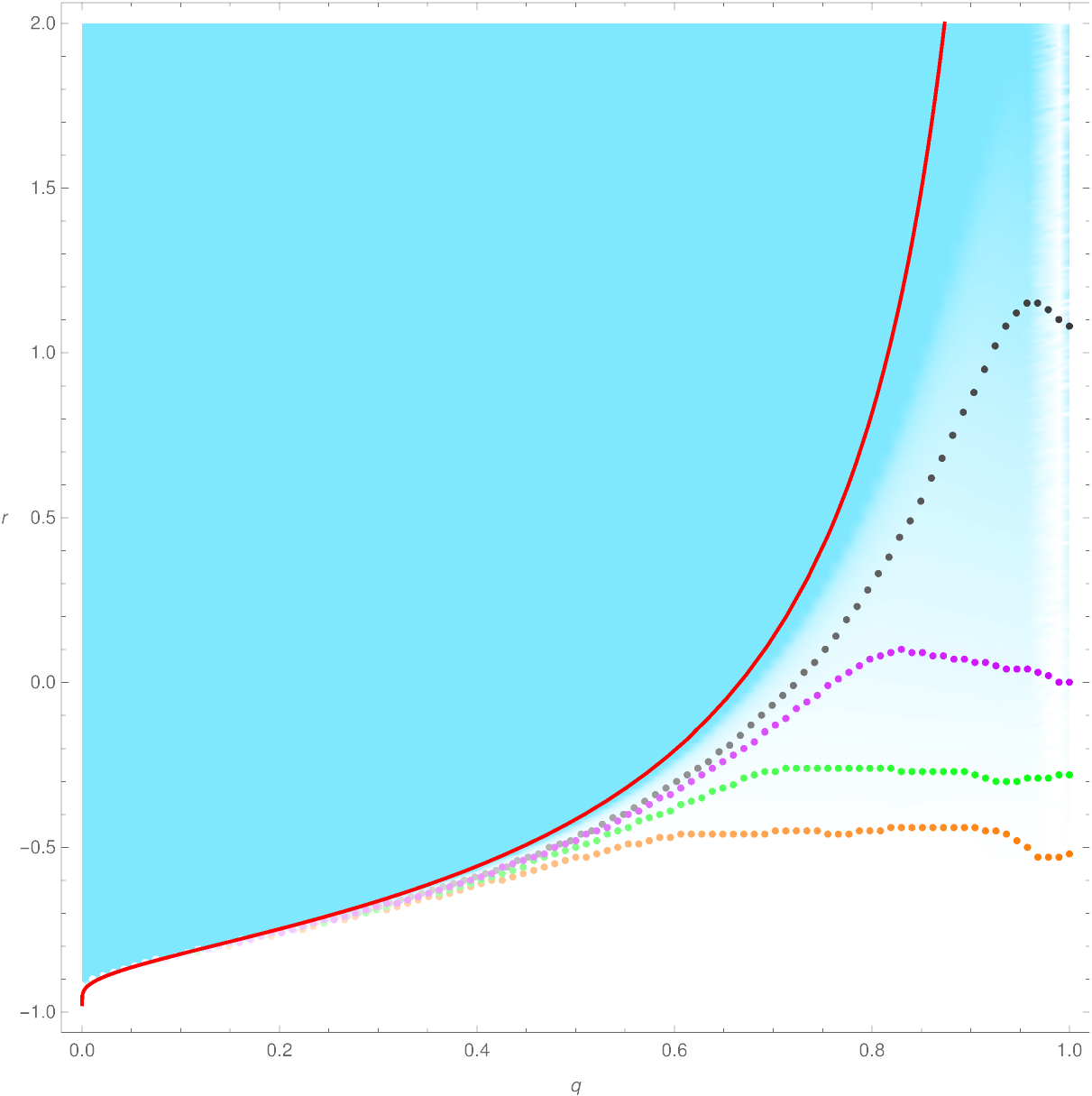}
        }
    \caption{The numerical analysis of the unitarity bound when $m^2=0$. We choose the spacetime dimension $4\leq d\leq 40$. In the left panel we use full partial-wave coefficients with $N\leq14$.  In the right panel we use partial-wave coefficients $c_{N,0}$ with $N\leq14$. The blue region satisfies partial wave unitarity, where the deeper blue represents the higher spacetime dimension $d$. We use the red line to denote the critical condition of $q$ and $r$ in \cref{eq:criticalcondition}, which matches the numerical results very well. We also label the special dimensions, including $d=4$ (orange), $6$ (green), $10$ (purple), and $26$ (black).}
    \label{fig:lowspinbound}
\end{figure}

We remind the readers that the hypergeometric Coon amplitude in four and six dimensions is not guaranteed to always be positive \cite{Cheung:2023adk}. But we emphasize that our study of these low spin partial-wave coefficients remains valuable. If we anticipate the existence of \textit{partial-wave low spin dominance}, we should expect that low spin data can fully control the unitarity bounds.

To boost our confidence in low spin dominance, we test the predictions for unitarity bounds from spin-0 data when $m^2=0$, see \cref{fig:lowspinbound}. Through numerical analysis of the partial-wave coefficients, we find that the unitarity bounds derived from spin-0 data align with those obtained from complete partial-wave coefficients. This alignment reinforces our belief in the existence of low spin dominance. We look forward to future work providing stronger evidence from both analytical and numerical perspectives.

\section{Various limits}\label{sec:Variouslimits}
The properties of the hypergeometric Coon amplitude allow us to easily investigate the behaviors across a broad class of amplitudes. As we delve into the context of the Veneziano amplitude, we discover that certain low spin coefficients at specific dimensions exhibit manifest positivity. We also test the \textit{partial-wave low spin dominance} within the superstring framework, providing potential evidence for manifest positivity below $d\leq10$. Finally, we revisit the Coon amplitude and re-derive the key results from the work by \cite{Rigatos:2024beq}.

\subsection{Veneziano amplitude}\label{sec:manifestpos}

Upon taking the limit as $q\to 1$ and $r\to 0$, the hypergeometric Coon amplitude reduces to the Veneziano amplitude. In this subsection, we focus on the superstring case where $m^2=0$. We will present the manifest positivity of superstring below $d\leq 6$ by using a novel integral representation at first. Next, we will demonstrate the existence of partial-wave low spin dominance and show the full analysis below $d\leq10$.

\subsubsection{Manifest Positivity of Superstring in $d\leq 6$}\label{sec:manifestpos}

We are going to following the conventions used in \cite{Arkani-Hamed:2022gsa}. Taking the limit as $q\to 1$, $r\to 0$, $m^2\to0$ and shifting on $N$, the partial-wave coefficients $c_{n,\ell}$ \cref{eq:res_doublehm} reduce to
\begin{align}\label{eq:res_superstringdoublehm}
     c_{N,\ell} =\sum_{n=0}^{N} \underbrace{\binom{n}{\ell} \frac{\sqrt{\pi} }{\mathcal{K}(\ell,\alpha)} 
     \frac{(-1)^{\ell} (\alpha)_{\frac{1}{2}+n}}{(\ell+2\alpha)_{1+n}}  }_{\mathcal{T}_{n,\ell}^{-1}}  \underbrace{ {\left(N+1\right)^{n-1}} \vphantom{\binom{a}{n}}}_{\text{scattering angle}} Z^{1,0}_{ {n}}(N)\; .
\end{align}
The tachyons in the super string is mass-less so there is a lack of mass term in this expression. Let us denote $Z^{1,0}_{ {n}}(N)$ by $Z_{n}(N)$ for convenience. The key observation is that these multiple harmonic numbers $Z_n(N)$ are generating by harmonic polylogarithms (HPL) \cite{Maitre:2005uu}.  Specifically, we have
\begin{equation}\label{eq:genehs}
    \frac{1}{1-z}H(1_1,1_2,\cdots,1_n;z)\equiv \frac{1}{1-z} \frac{(-1)^n}{n!} \log^n(1-z)=\sum_{N=0}^\infty Z_{n}(N)z^N\; .
\end{equation}
By substituting \cref{eq:res_superstringdoublehm} into \cref{eq:genehs}, we can extend the upper bound of $n$ to $\infty$ and complete the summation. This allow us to express the partial-wave coefficients $c_{N,\ell}$ as the series coefficients of a specific function, yielding
\begin{align}\label{eq:superstringser}
    \frac{2(-1)^{\ell}}{(1-z)} \frac{\ell+\alpha}{N+1} \Gamma(2\alpha)  _2\tilde{F}_2 \left(1,\alpha+\frac{1}{2};1-\ell, \ell+2\alpha+1;(N+1)\log(1-z) \right) \sim c_{N,\ell} z^{N} \, .
\end{align}
We can extract the $c_{N,\ell}$ using the residue theorem. Packaging the series coefficients, we obtain
\begin{align}\label{eq:superstringint}
    c&_{N,\ell} =\frac{1}{2\pi i} \oint \mathrm{d}z\frac{2(-1)^{\ell}}{z^{N+1}(1-z)} \frac{\ell+\alpha}{N+1} \Gamma(2\alpha)  _2\tilde{F}_2 \left(1,\alpha+\frac{1}{2};1-\ell, \ell+2\alpha+1;(N+1)\log(1-z) \right)\, .
\end{align}
This integral form involves all information in a closed form. It also provide a feasible path to present the manifest positivity of super string.

\begin{table}[h!]
\begin{center}
\begin{tabular}{c|c c c c c}
Veneziano & $\ell=0$ & $\ell=1$ & $\ell=2$ &$\ell=3$ & $\ell=4$ \\ \hline
$N=0$ & $ 1 $ &  &  &  &\\
$N=1$ & $ 0 $ & $\frac{1}{2(d-3)}$  &  &\\
$N=2$ & $\frac{10-d}{24(d-1)}$ &  $ 0 $ & $ \frac{3}{4 (d-1) (d-3)} $ & & \\
$N=3$ & $ 0 $ &  $ \frac{11-d}{12 (d-3) (d+1)} $ & $ 0 $ & $ \frac{2}{(d-3) (d^2-1)} $ &  \\
$N=4$ & $ \frac{9 d^2-250 d+1616}{1920 (d-1) (d+1)} $ &  $ 0 $ & $ \frac{25 (12-d)}{96 (d-1) (d^2-9) } $ & $ 0 $ & $\frac{125}{16  (d^2-1) (d^2-9)}$ \\
\end{tabular}
\end{center}
\caption{Partial-wave coefficients of super string for low-lying $(N,\ell)$. These results 
are in agreement with \cite[Table 1]{Arkani-Hamed:2022gsa}. Note that $N=0,1,2,3,\cdots$ hence there is a shift in $N$ comparing with \cite{Arkani-Hamed:2022gsa}.}
\label{Table:superstring}
\end{table}

Before we proceed, let us generate some data to check the consistency of our results. We list several partial-wave coefficients from \cref{eq:res_superstringdoublehm,eq:superstringser} in \cref{Table:superstring}, which matches \cite[Table 1]{Arkani-Hamed:2022gsa} precisely. \footnote{Compared to \cite{Maity:2021obe,Rigatos:2023asb} there is a shift in the quantum numbers and this is why the coefficients $c_{N,\ell}$ for $N+\ell$ even are non-zero.}

\begin{figure}[t]
    \centering
    \begin{tikzpicture}[>={Kite[inset=0pt,length=0.3cm,bend]},
      decoration={markings,
      mark= at position 0.15 with {\arrow{>}},
      mark= at position 0.42 with {\arrow{>}},
      mark= at position 0.65 with {\arrow{>}},
      mark= at position 0.82 with {\arrow{>}},
      mark= at position 0.95 with {\arrow{>}},
      },scale=0.75]

\def\gap{0.25}
\def\bigradius{3.5}
\def\littleradius{0.8}
\def\lradius{0.3}

\draw[black, thick,   decoration, 
      postaction={decorate}]  
  let
     \n1 = {asin(\gap/2/\bigradius)},
     \n2 = {asin(\gap/2/\littleradius)}
  in (\n1:\bigradius) arc (\n1:360-\n1:\bigradius) 
  -- (-\n2:\littleradius) arc (-\n2:0+\n2:\littleradius)
  -- cycle;

\draw[black, thick, ->] 
      (\lradius,0) arc (0:360:\lradius);

\node[above left] at (-0.1,0.1) {$C_1$};
\node[below right] at (-2.5,2.5) {$C_{\infty}$};
\node[above] at (2.5,0.2) {$C_+$};
\node[below] at (2.5,-0.2) {$C_-$};
\node[below] at (-\bigradius+0.5,0) {$R$};

\node[] at (\littleradius+0.1,0) {$\times$};
\draw[snake=bumps,color=red] (\littleradius+0.1,-0.05) -- (1.1*\bigradius,-0.05);

\draw[fill=black] (0,0) circle(0.08);

\draw[-Latex] (-1.2*\bigradius,0) -- (1.2*\bigradius,0) node[below]{$\Re(z)$} ;
\draw[-Latex] (0,-1.1*\bigradius) -- (0,1.2*\bigradius) node[right]{$\Im(z)$};

\end{tikzpicture}
    \caption{Contours in the complex plane. We present the initial contour $C_1$ and the deformed contour $C_\infty+C_-+C_+$. We also use red bumps to label the branch cut from $\log(1-z)$.}
    \label{fig:contour}
\end{figure}
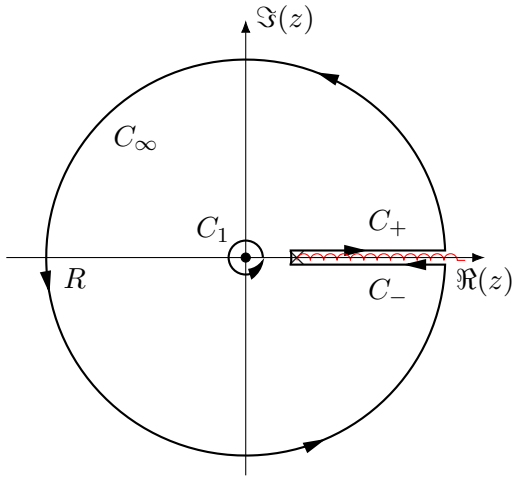

Now, let us formally investigate closely the concept of manifest positivity within superstring theory. Starting from \cref{eq:superstringint}, we observe that the integrand has only a single branch cut, commencing at $z=1$. We can perform a contour deformation, as depicted in \cref{fig:contour}. Remarkably, this contour deformation effectively yields the integral of the Disc of the integrand in \cref{eq:superstringint}.

For our analysis, we firstly set $d=4$ and $\ell=0$ in \cref{eq:superstringint}. This choice leads us to the expression
\begin{equation}
    c^{(4)}_{N,0}=\frac{1}{(N+1)^2}\int_{1}^{\infty}\frac{\mathrm{d}z}{z^{N+1}}\frac{\left(1-(1-z)^{N+1}\right) }{ (z-1) \left(\log ^2(z-1)+\pi ^2\right)} \; ,
\end{equation}
When $N$ takes the odd values the partial-wave coefficients yield zero. Therefore, we need focus on cases where $N$ is even. Given that $z\geq1$, we conclude this expression remains \textit{always positive}.

The coefficient $c^{(4)}_{N,1}$ warrants a more detailed analysis, although the fundamental essence remains similar. Its integral expression is given by
\begin{equation}
    c^{(4)}_{N,1}=\frac{3}{(N+1)^3} \int_{1}^{\infty}\frac{\mathrm{d}z}{z^{N+1}}\frac{F(z)+G(z)}{(z-1) \left(\log ^2(z-1)+\pi ^2\right)^2} \; ,
\end{equation}
For ease of analysis, we decompose the integrand into two parts, denoted by
\begin{align}
    F(z)= & (1-z)^{N+1} \log (z-1) ((N+1) \log (z-1)-4)+N \pi ^2 \; ,\\
    G(z)= & \pi ^2 (N+1) (1-z)^{N+1}+(N+1) \log ^2(z-1)+4 \log (z-1)+\pi ^2 \; ,
\end{align}
Now, let us focus solely on the cases where $N$ is odd. We begin by examining the simpler function $F(z)$, where we can prove that its function values at the stationary are consistently positive. Notably, the stationary of $F(z)$ and function values are given by
\begin{equation}\label{eq:c41F}
    z ^* = 1+\exp \left(\frac{1 \pm \sqrt{5}}{N+1}\right), \quad F(z^*)=\pi ^2 N+\frac{2 e^{\sqrt{5}+1} \left(1\mp\sqrt{5}\right)}{N+1} >0 ,\quad \forall \;  N>1 \; .
\end{equation}
Additionally, we have $F(1)=(N+1) \pi^2$ and $F(\infty)\to \infty$. Hence we establish that $F(z)\geq0$ when $z\geq1$.

Next we turn to $G(z)$. The logic for proving its positivity is slightly more intricate. Instead of focusing on the function values at the stationary point $z^*$, we consider the entire function on the interval $z^*\geq1$. Thus, we have:
\begin{equation}\label{eq:c41G}
    G(z)\geq G(z^*)=\log (z^*-1) ((N+1) \log (z^*-1)+2)-\frac{4}{N+1}+\pi ^2 \geq \pi ^2-\frac{5}{N+1} \; ,
\end{equation}
where $z^*$ represents the stationary point of $G(z)$. In conclusion, combining \cref{eq:c41F,eq:c41G}, we demonstrate that the partial-wave coefficient $c^{(4)}_{N,1}$ is \textit{manifestly positive}.

\begin{figure}[t!]
    \centering
    \subfloat[$d=4$, $\ell=0$]{
        \includegraphics[width=5cm]{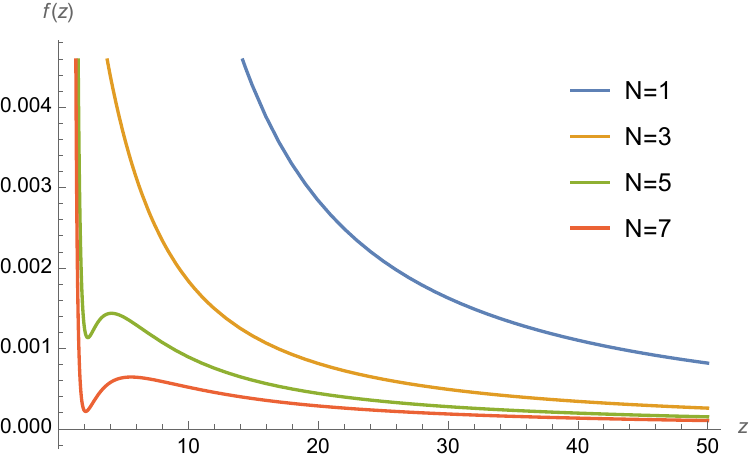}
        }
    \subfloat[$d=4$, $\ell=1$]{
        \includegraphics[width=5cm]{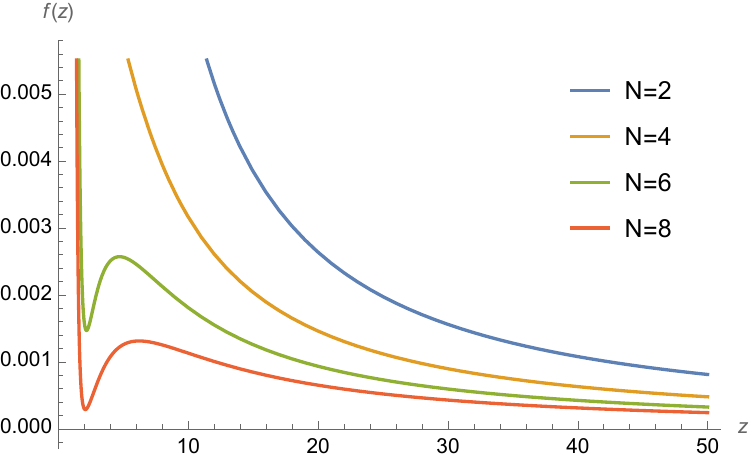}
        }
    \subfloat[$d=4$, $\ell=2$]{
        \includegraphics[width=5cm]{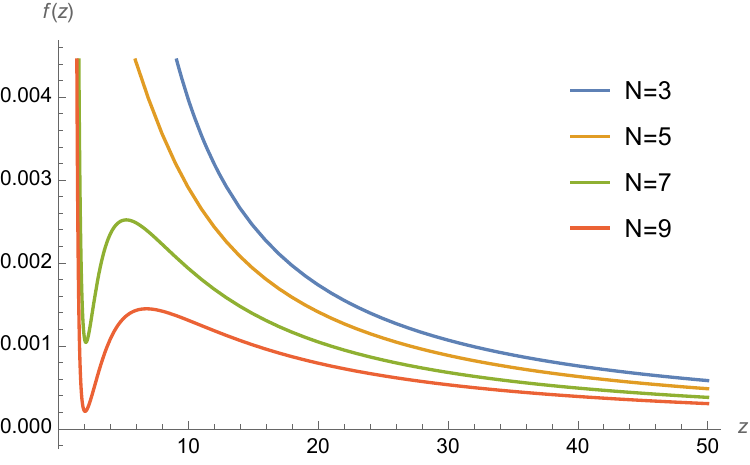}
        }\\
    \subfloat[$d=6$, $\ell=0$]{
        \includegraphics[width=5cm]{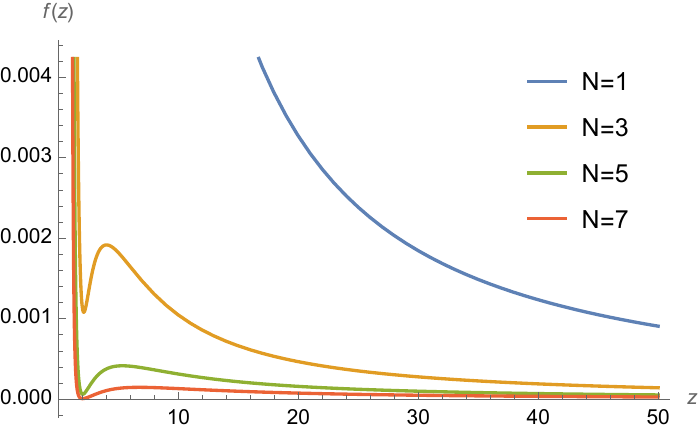}
        }
    \subfloat[$d=6$, $\ell=1$]{
        \includegraphics[width=5cm]{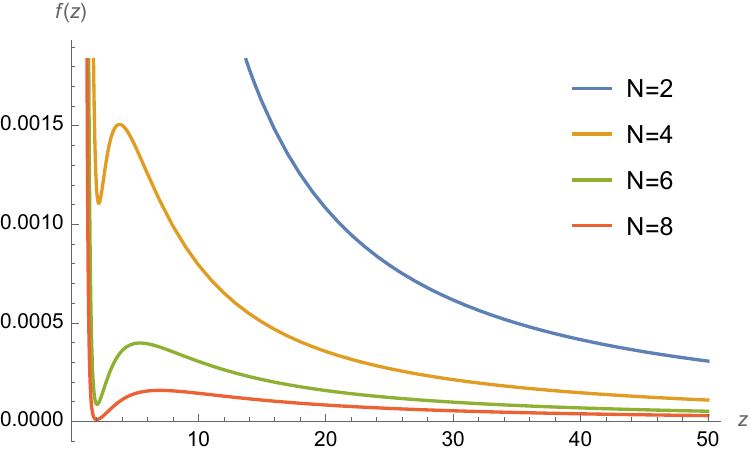}
        }
    \subfloat[$d=6$, $\ell=2$]{
        \includegraphics[width=5cm]{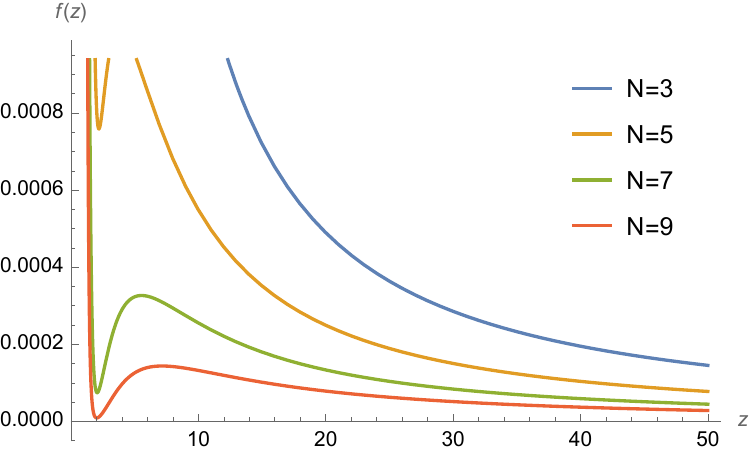}
        }
    \caption{The integrand in \cref{eq:superstringint} after analytic continuation as \cref{fig:contour}. We present some low-lying data upon $d\leq6$ and $\ell\leq2$.}
    \label{fig:genepartialwave}
\end{figure}

We are led to examine a more complicated coefficient $c^{(6)}_{N,0}$. As before, we begin by writing down the integral expression
\begin{equation}
    c^{(6)}_{N,0}=\frac{12}{(N+1)^4} \int_{1}^{\infty}\frac{\mathrm{d}z}{z^{N+1}} \frac{\pi ^2 \left(1-(1-z)^{N}\right) -\log (z-1) H(z) }{ (z-1) \left(\log ^2(z-1)+\pi ^2\right)^3} \; .
\end{equation}
For our analysis, we focus on cases where $N$ is even. Let us point out that $\pi ^2 \left(1-(1-z)^{N}\right)$ is always positive at a very beginning. Then we examine the part involving $-log(z-1)H(z)$. Specifically, we have
\begin{equation}
     H(z)=(N+1) \left((1-z)^{N+1}+1\right) \left(\log ^2(z-1)+\pi ^2\right)+3 \left(1-(1-z)^{N+1}\right) \log (z-1)\; .
\end{equation}
Upon observing that $H(z)$ is monotonically decreasing and possesses only one zero at $H(2)=0$, we conclude that $-log(z-1)H(z)$ is non-negative. Consequently, we establish that $c^{(6)}_{N,0}$ is also \textit{manifestly positive}.

The discussion regarding specific examples concludes here. We present more aspects of the general partial-wave coefficients for dimensions $d=4$ and $d=6$ in \cref{fig:genepartialwave}. Let us emphasize that the deformed integrand in \cref{eq:superstringser} is not always positive. Specifically, for a fixed spin $\ell$, this integrand only exhibits manifest positivity when $N\geq3\ell-2d+1$. However, we believe that the approach using the generating function of harmonic numbers still remains worthwhile for future purposes.

In fact, \cref{eq:genehs} is not an isolated case but represents a more general situation. Starting from \cref{eq:genehs} we can derive a family of functions, each possessing an integral form similar to that in \cref{eq:superstringint}. The open question remains: Can we ingeniously combine these functions and establish the manifest positivity of super string below $d\leq10$?

\subsubsection{Partial-wave low spin dominance in Superstring}

We aim to partially answer this question through the potential existence of low spin dominance. Through our study on the hypergeometric Coon amplitude, we discovered that low spin datas effectively control the unitarity bound. A similar phenomenon exists for the super string amplitude. 

\begin{figure}[t]
    \centering
    \subfloat[Critical dimensions $d^*$ in Regge trajectories up to 21-th.]{
        \includegraphics[width=7cm]{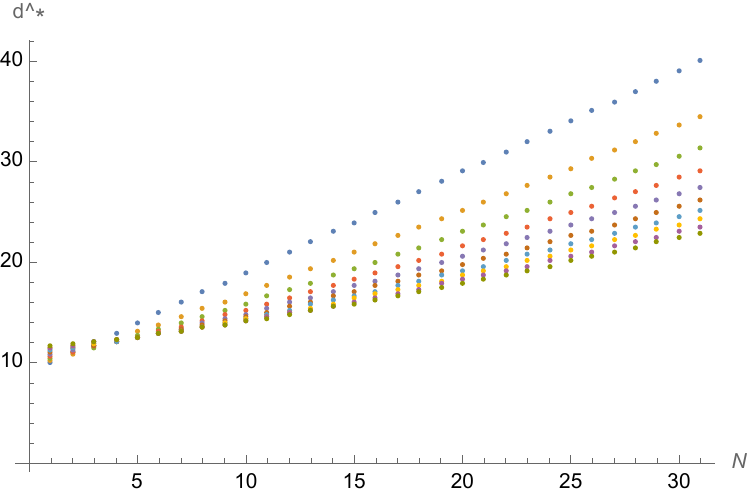}
        } \hfill
    \subfloat[Critical dimensions $d^*$ in special Regge trajectories: 31-th(blue), 51-th(yellow) and 71-th(green).]{
        \includegraphics[width=7cm]{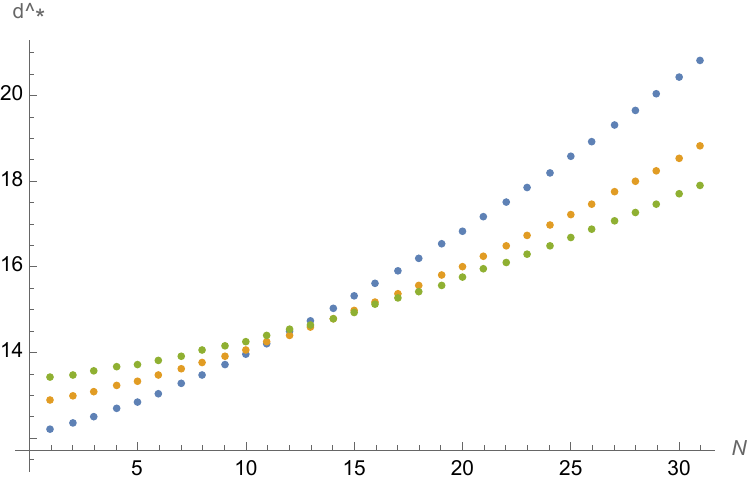}
        }
    \caption{The numerical analysis of the critical dimension $d^*$ in super string. The same color correspondence to $d^*$ in the same Regge trajectory.}
    \label{fig:Reggebehivor}
\end{figure}

Let us consider a Regge trajectory $c_{N,N-2}$. The behavior of these partial-wave coefficients is given by
\begin{equation}
    c_{N,N-2} \propto \frac{2 \alpha -N-5}{(\alpha +N-1) (\alpha )_{N-2}} \; .
\end{equation}
We can directly solve the phase transition dimension. The critical dimension $d^*$ is
\begin{equation}
    d^*=N+8 \; .
\end{equation}
The critical dimension of the super string $d=10$ is given by the case $N=2$; when $N>2$, we find that all bounds are covered by the coefficient $c_{2,0}$ of spin $0$. This inspires us to study this phenomenon more meticulously. We have examined all Regge trajectories up to $N-\ell=71$, and we found that within the same trajectory, the entire critical dimension is always controlled by the data of spin $0$, as shown in \cref{fig:Reggebehivor}. Combining the low spin dominance and positivity in $d = 10$ for low spin, we can demonstrate the unitarity in $d=10$ manifestly.

The partial-wave coefficients $c_{N,0}$ in $d= 10$ has been proven to be always positive in \cite{Arkani-Hamed:2022gsa}. We also provide another brief argument for this manifest positivity. Let us start from
\begin{align}
    \int_1^{z} \mathrm{d}t \frac{2(-1)^{\ell}}{(1-z)} \frac{\ell+\alpha}{N+1} \Gamma(2\alpha)  _2\tilde{F}_2 \left(1,\alpha+\frac{1}{2};1-\ell, \ell+2\alpha+1;(N+1)\log(1-t) \right) \sim \frac{c_{N,\ell}}{N+1} z^{N+1} \, .
\end{align}
Considering $d=10$ and $\ell=0$, this partial-wave coefficient can be written as
\begin{equation}\label{eq:superd10l0}
    c^{(10)}_{N,0}=\frac{280}{(N+1)^7} \frac{1}{2\pi i}\oint \frac{\mathrm{d}z}{z^{N+2}} \left(I_1(z)+I_2(z)\right)\; , 
\end{equation}
where
\begin{align}
    I_1(z)=& \frac{60-60 (1-z)^{N+1}}{\log ^6(1-z)}+\frac{24 (N+1) (1-z)^{N+1}+36 (N+1)}{\log ^5(1-z)}\; \\
    I_2(z)=&\frac{9 (N+1)^2-3 (N+1)^2 (1-z)^{N+1}}{\log ^4(1-z)}+\frac{(N+1)^3}{\log ^3(1-z)}\; .
\end{align}
Focusing on $N$ is even we observe that the series coefficients of $I_1(z)$ and $I_2(z)$ can be bounded by linear functions of $N$, which leads us to
\begin{equation}\label{eq:boundd10l0}
    c^{(10)}_{N,0} \geq \frac{280}{(N+1)^7} \frac{1} {2\pi i} \oint \frac{\mathrm{d}z}{z} \left(\left(-\frac{3 N}{4000}-\frac{1}{25} \right)+\left(\frac{N}{2}-25\right) \right)\; .
\end{equation}
The RHS of \cref{eq:boundd10l0} is \textit{manifestly positive} when $N>50$, and it is straightforward to check that $c^{(10)}_{N,0}$ in \cref{eq:superd10l0} is non-negative when $N\leq50$. This method can be extended to higher spins, but we leave it to the future research.

By now we can summarize the strategy of our argument 
\begin{itemize}
    \item The partial-wave coefficients $c^{(10)}_{N,0}$ is manifest positive and the critical dimension $d^*=10$ is given by $c^{(4)}_{2,0}$.
    \item The critical dimensions determined by the specific Regge trajectory are controlled by the spin-$0$ data.
\end{itemize}
In conclusion, we propose a new insight that the super string is manifestly positive below $d\leq10$. The central concept is that by observing certain fixed Regge trajectories, we recognize that the coefficients of spin-$0$ efficiently determine the unitary bound of the entire trajectory. We are currently unable to complete the proof analytically, but numerical analysis provides suggestive evidence. We anticipate a comprehensive understanding of existence of low spin dominance in the partial-wave unitarity analysis.

\subsection{Coon amplitude}
In this subsection, we will succinctly revisit the main result of the partial-wave analysis of the Coon amplitude. Many of these results can be directly derived from the hypergeometric Coon results, and some analyses have been meticulously examined in our previous work \cite{Rigatos:2024beq}. Therefore, we will briefly recapitulate the crucial conclusions here.

We can derive the partial-wave coefficients of Coon amplitude directly from the hypergeometric Coon results \cref{eq:res_doublehm}, which reads
\begin{align}\label{eq:coonres_doublehm}
     c_{N,\ell} ={q^N} \sum_{n,k=0}^{N} \underbrace{\binom{n}{\ell} \frac{\sqrt{\pi} }{\mathcal{K}(\ell,\alpha)} 
     \frac{(-1)^{\ell} (\alpha)_{\frac{1}{2}+n}}{(\ell+2\alpha)_{1+n}}  }_{\mathcal{T}_{n,\ell}^{-1}}  \underbrace{\binom{k}{n} (-m^2)^{k-n}}_{\text{external mass}} \underbrace{ {\left(3 m^2-[N]_q\right)^n} \vphantom{\binom{a}{n}}}_{\text{scattering angle}} Z^{q,0}_{ {k}}(N)\; .
\end{align}
Obviously, our method is applicable straightforwardly in the Coon amplitude. The harmonic numbers $Z^{q,0}_k(N)$ are also called multiple harmonic $q$-series in \cite{Bradley_2005}. 

By using the symptotic behavior of harmonic numbers $Z^{q,0}_k(N)$ we can derive the
\begin{align}
    \lim_{N\to\infty }c_{N,0}&\propto \lim_{N\to\infty } \left[  q^N-\frac{N}{2}  q^N \left(-3 m^2 +[N]_q\right)  Z^{q,0}_1(N) +\cdots \right] \nonumber \\
    &\propto2 q^{-1}-m^2 (q-1)-3\; .
\end{align}
Then we can solve the critical value of $q_\infty(m^2)$, where
\begin{equation}
    q_{\infty}(m^2)=\frac{m^2-3+\sqrt{9+2m^2+m^4}}{2m^2}\; .
\end{equation}
We conclude that the Coon amplitude is manifestly positive when $q\leq q_\infty(m^2)$ in any spacetime dimensions $d$.

The Regge trajectory analysis of Coon amplitude brings unitarity bounds $-1\leq m^2\leq\frac{1}{3}$ and shows the manifest positivity in the sub-leading trajectory.  The utilization of harmonic numbers benefits the analysis of Regge trajectories.

Furthermore, our new approach is efficacious in the low-spin analysis. We present the low spin partial-wave coefficients in a simple form. From numerical analysis of the partial-wave coefficients in hypergeometric Coon amplitude, we find some hint of the existence of low spin dominance. In the case of massless Coon amplitude, the unitarity bounds derived from spin-0 data are in agreement with those obtained from the complete partial-wave coefficients.

\section{Discussion}\label{sec:outlook}

In this work, based on harmonic numbers, we have introduced a novel method to deal with the partial-wave unitarity of the hypergeometric Coon amplitude. This can be seen as a follow-up of \cite{Rigatos:2024beq} and sufficiently demonstrates that this technique can be straightforwardly applied to the hypergeometric Coon amplitude, even though it is a highly non-trivial deformed amplitude. Hence, this provides even more concrete evidence that the basis of harmonic numbers in the partial-wave unitarity analysis is generalizable in a straightforward manner for any amplitude.

Our new approach derived some unitarity bounds. The Regge trajectory analysis provide the bounds \cref{eq:unitaritylead,eq:unitarysubleading}. The asymptotic behavior of harmonic numbers yield the critical condition of $q$ and $r$ \cref{eq:criticalcondition}. The simplification appears in low-spin partial-wave coefficients \cref{eq:c4N0,eq:c4N1,eq:c6N0}. These low-spin data describe the quantitative behavior of the whole unitarity analysis very well.

We also discussed the manifest positivity of super string in various dimensions. By substituting the generating function of ordinary multiple harmonic numbers $Z_n(N)$ \cref{eq:genehs}, we present the manifest positivity of super string in below $d=6$. Combining the low-spin data $c^{(10)}_{N,0}$ is manifest positive and partial-wave low spin dominance, we  demonstrate a new insight on the manifest positivity of super string in $d=10$.

Let us revisit the merits of our novel method. It primarily offers a suggestive and well-structured form of the final answer. It yields a way to solve the coefficients by reducing the task into a linear algebra problem. A third advantage is the utilization of harmonic numbers and their deformations, which facilitates the analysis of non-trivially deformed amplitudes in a straightforward manner. Moreover, owing to the properties of the harmonic numbers, the simplified cases of the deformations are inherited. Lastly, this new approach simplifies the low-spin trajectory analysis. In conclusion, we recommend that the results obtained using harmonic numbers remain valuable for both analytic and numerical analysis.

Our methodology can be readily adaptable to a variety of scenarios. It proficiently handles the intricate and more extensively deformed amplitude: the hypergeometric Coon amplitude. This amplitude approaches broad known amplitudes in different limits. We suggest that harmonic numbers also furnish an application for the partial-wave analysis of the Vrasoro-Shapiro amplitude \cite{Virasoro:1969me,Shapiro:1970gy} and its hypergeometric generalisation \cite{Cheung:2023adk}. It is also interesting to consider if we can use harmonic numbers in more general Regge trajectories \cite{Eckner:2024ggx} and conformal background \cite{vanRees:2023fcf}.

The algebra properties of harmonic numbers are worth to be explored. The shuffle algebra between harmonic polylogarithms (HPL)\cite{Maitre:2005uu} provide a nice identity \cref{eq:genehs}.The $q$-deformed multiple polylogarithms \cite{schlesinger2001remarks} may furnish the similar identity and simplify the analysis.

Beyond partial-wave unitarity analysis, harmonic numbers can be utilized in the AdS Virasoro-Shapiro amplitude as a nice basis \cite{Alday:2022uxp,Alday:2022xwz}. Furthermore, a  potential application could be to provide additional information in the  calculation of the Wilson coefficients \cite{Haring:2023zwu}.

Motivated by recent works on the S-matrix bootstrap, an interesting future study would be to consider where these new amplitudes are in the EFT-hedron \cite{Arkani-Hamed:2020blm}; do they saturate any surface? It becomes imperative to employ innovative methodologies to validate the physicality of these new amplitudes.

Finally, the Veneziano amplitude and the Virasoro-Shapiro amplitude are related by the KLT relation \cite{Kawai:1985xq}. It would be interesting to investigate whether an analogous relation is applicable for the hypergeometric deformations of the Veneziano and Virasoro-Shapiro amplitudes, thereby leading to a hypergeometric KLT relation.

~\\

\noindent{\bf Acknowledgements.} BW would like to thank Zhongjie Huang, Konstantinos C. Rigatos and Ellis Ye Yuan for helpful discussions. BW is also grateful to Konstantinos C. Rigatos and Ellis Ye Yuan for the detailed comments on the draft. The work of BW is supported by National Science Foundation of China under Grant No.~12175197, Grant No.~11935013 and Grand No.~12347103. BW is also supported by the Fundamental Research Funds for the Chinese Central Universities under Grant No.~226-2022-00216.

\bibliography{aps}
\bibliographystyle{JHEP}

\end{document}